\documentclass[aps,prl,amssymb,superscriptaddress,footinbib,twocolumn]{revtex4} 
\usepackage{graphicx}
\usepackage{amsmath}
\usepackage{color}
\usepackage{cancel}
\usepackage{ulem} 
\usepackage{marginnote}
\usepackage{tikz}

\begin{document}

\title{Quantum discord in the Dynamical Casimir Effect}

 \author{Carlos Sab\'in}
\affiliation{School of Mathematical Sciences, University of Nottingham,
  NG7 2RD Nottingham, United Kingdom}
\email{carlos.sabin@nottingham.ac.uk}
\author{Ivette Fuentes}
\affiliation{University of Vienna, Faculty of Physics, Boltzmanngasse 5, 1090 Wien, Austria}\affiliation{School of Mathematical Sciences, University of Nottingham,
  NG7 2RD Nottingham, United Kingdom}
\author{G\"oran Johansson}
\affiliation{Microtechnology and Nanoscience, MC2, Chalmers University of Technology, SE-412 96 G{\"o}teborg, Sweden}

  \date{\today}

\begin{abstract}
We analyse the generation of quantum discord by means of the dynamical Casimir effect in superconducting waveguides modulated by superconducting quantum interferometric devices. We show that for realistic experimental parameters, the conditions for the existence of quantum discord are less demanding than the previously considered for quantum entanglement or non-classicality. These results could facilitate the experimental confirmation of the quantum nature of the dynamical Casimir effect radiation. Moreover, the states with non-zero discord and zero entanglement generated by the dynamical Casimir effect are a useful resource for quantum cryptography.

\end{abstract}
\maketitle

The Dynamical Casimir Effect (DCE) consists in the generation of photons out of the vacuum of a quantum field, by means of the modulation of a boundary condition -e. g. a mirror- at relativistic speeds. The phenomenon  was theoretically predicted in 1970 \cite{moore}. However, it was not until 2011 that the first experimental observation was reported \cite{casimirwilson}. The reason for this is that accelerating a mirror to relativistic speeds  by mechanical means is out of experimental reach. The difficulty was circumvented by implementing a different kind of mirror. In particular,  a Superconducting Quantum Interferometric Device (SQUID) which interrupts a superconducting transmission line provides a boundary condition equivalent to a mirror under some experimental conditions. Unlike its mechanical counterpart this boundary condition can be made to move at velocities close to the speed of light in the medium, enabling the observation of radiation generated by the DCE.

A distinctive feature of the DCE radiation is its genuinely quantum nature, due to its origin in the virtual particles of the quantum vacuum. In particular, the collected radiation should display entanglement, in stark distinction to the one originated from the modulation of a thermal source. However, the experimental confirmation of the existence of DCE entanglement has proven difficult so far. In \cite{nonclassicaldce}, it is shown that entanglement shows up above a critical threshold of thermal noise and the conditions for the experimental amplitudes and velocities of the boundary condition necessary to overcome that noise are provided. Experimental efforts towards it are currently going on in the laboratory. In this work, we analyse an alternative strategy to confirm the quantum nature of the DCE radiation. We consider the generation of a more general form of quantum correlations, such as quantum discord.

Quantum discord \cite{OZ01,OZ02,HV01} has attracted a great deal of attention in the last years and represents a paradigm shift in the analysis of quantum correlations. In particular, it has been shown that some quantum states that do not possess entanglement still display some genuinely quantum form of correlations -characterised by the discord- that are useful for quantum technologies, such as remote state preparation \cite{discordtech1, discordtech2}, quantum metrology \cite{discordtech3}, microwave quantum illumination \cite{barjanzeh} and quantum cryptography \cite{discordcryp}.

In this paper, we compute the quantum discord generated in the experimental setup employed for the observation of the DCE, namely a superconducting waveguide interrupted by a SQUID. Using realistic experimental parameters, we show that the driving amplitudes and velocities that are required to overcome a certain level of thermal noise and generate quantum discord are smaller than the ones necessary to generate quantum entanglement and other nonclassicality indicators discussed in \cite{nonclassicaldce}. In other words, we show that for a given experimental value of the driving amplitude and velocity, the critical value of temperature above which quantum discord vanishes is higher than in the case of quantum entanglement. These results could facilitate the experimental confirmation of the quantum nature of DCE radiation. Moreover,  the kind of states with non-zero discord and zero entanglement generated by the DCE in our setup have proven to be a useful resource in device-dependent continuous variable quantum cryptography \cite{discordcryp}.

Let us now discuss our model and results in detail. We will consider the same experimental setup as in \cite{casimirwilson, nonclassicaldce}. The electromagnetic field confined by a superconducting waveguide is described by a quantum field associated to the flux operator $\Phi(x,t)$., 
which obeys the 1+1 D Klein-Gordon wave equation, $\partial_{xx}\Phi(x,t)-v^{-2}\partial_{tt}\Phi(x,t)=0$. The  field can thus be written in the form
\begin{eqnarray}
\label{eq:field}
\Phi(x,t) &=& \sqrt{\frac{\hbar Z_0}{4\pi}}\int_{-\infty}^{\infty} \frac{d\omega}{\sqrt{|\omega|}}\times\nonumber\\
&&
\left[a(\omega) e^{-i(-k_\omega x +\omega t)} + b(\omega)e^{-i(k_\omega x +\omega t)}\right],
\end{eqnarray}
where $a(\omega)$ and $b(\omega)$ are the annihilation operators for photons with frequency $\omega$ propagating to the right -incoming- and left -outgoing-, respectively. Here we have used the notation $a(-\omega)=a^\dag(\omega)$, and $k_\omega = \omega/v$ is the wavenumber, $v$ is the speed of light in the waveguide, and $Z_0$ the characteristic impedance.

As shown in \cite{johansson:2009,johansson:2010}, for large enough SQUID plasma frequency, the SQUID provides the following boundary condition to the flux field:
\begin{eqnarray}
\Phi(0,t) + \left.L_{\rm eff}(t)\partial_x\Phi(x,t)\right|_{x=0} = 0,
\end{eqnarray}
that can be described by an effective length 
\begin{equation}
L_{\rm eff}(t) = \left(\Phi_0/2\pi\right)^2/(E_J(t)L_0), 
\end{equation}
where $L_0$ is the characteristic inductance per unit length of the waveguide and $E_J(t)=E_J[\Phi_{\rm ext}(t)]$ is the flux-dependent effective Josephson energy.  For sinusoidal modulation with driving frequency $\omega_d/2\pi$ and normalized amplitude $\epsilon$, $E_J(t) = E_J^0 [1 + \epsilon \sin \omega_d t]$, we obtain an effective length modulation amplitude $\delta\!L_{\rm eff} = \epsilon L^0_{\rm eff}$, where $L^0_{\rm eff} = L_{\rm eff}(0)$. If the effective velocity $v_{\rm eff}=\delta\!L_{\rm eff}\omega_d$ is large enough as compared to $v$, the emission of photons by means of the DCE is sizeable.

Within this framework, the DCE is analysed using scattering theory which describes how the time-dependent boundary condition mixes the otherwise independent incoming and outgoing modes \cite{lambrecht:1996}. 
In the perturbative regime discussed analytically in \cite{johansson:2009,johansson:2010,nonclassicaldce}, the resulting output field is correlated at modes with angular frequencies $\omega_+, \omega_-$, such that $\omega_+ + \omega_-=\omega_d$, so we can write $\omega_\pm = \omega_d/2 \pm \delta\omega$, where $\delta\omega$ is the detuning.
Introducing the notation $a_\pm=a(\omega_\pm)$ and $b_\pm=b(\omega_\pm)$, the relation between the input and the output fields is the following:
\begin{eqnarray}
\label{eq:output-field-perturbation-simplified-notation}
b_\pm = -a_\pm -i\frac{\delta\!L_{\rm eff}}{v}\sqrt{\omega_+\omega_-}a^\dag_\mp,
\end{eqnarray}
where  $\delta\!L_{\rm eff}\sqrt{\omega_-\omega_+}/v$ is a small parameter. If we consider small detuning, then $\omega_-\simeq \omega_+\simeq\omega_d/2$ and
\begin{equation}
\frac{\delta\!L_{\rm eff}\sqrt{\omega_-\omega_+}}{v}\simeq\frac{\epsilon L_{\rm eff}(0)\omega_d}{2v}=\frac{v_{\rm{eff}}}{2v}. 
\end{equation}
Denoting the small parameter as $f$, we can write:
\begin{eqnarray}
\label{eq:output-field-perturbation-simplified-notation2}
b_\pm = -a_\pm -i\,f\,a^{\dag}_\mp.
\end{eqnarray}

Let us consider now the covariance matrix of the system $V$. Using the same convention as in \cite{nonclassicaldce} $$V_{\alpha\beta} = \frac{1}{2}\left<R_\alpha R_\beta+R_\beta R_\alpha\right>,$$ -which assumes zero displacement- where $$R^{\rm T} = \left(q_-, p_-, q_+, p_+\right)$$ is a vector with the quadratures as elements: $$q_\pm = (b_\pm + b_\pm^\dag)/\sqrt{2}$$ and $$p_\pm = -i(b_\pm - b_\pm^\dag)/\sqrt{2}.$$ Note that the quadratures of the outgoing modes can be written in terms of the ingoing modes $$q_{0\pm} = (a_\pm + a_\pm^\dag)/\sqrt{2}$$ and $$p_{0\pm} = -i(a_\pm - a_\pm^\dag)/\sqrt{2}$$ by using Eq. (\ref{eq:output-field-perturbation-simplified-notation2}):
\begin{eqnarray}
\label{eq:cuadratures}
q_\pm=-(q_{0\pm}+\,f\,p_{0\mp})\nonumber\\
p_\pm=-(p_{0\pm}+\,f\,q_{0\mp}).
\end{eqnarray}
We assume that the ingoing modes are in a weakly thermal, quasi-vacuum state characterised by a small fraction of thermal photons $n^{\rm{th}}_+$, $n^{\rm{th}}_-$ as is the case for typical $\rm{GHz}$ frequencies and $\rm{mK}$ temperatures  in a superconducting scenario. Then the ingoing covariance matrix is 
\begin{equation}
\label{eq:covariancematrixin}
V_0=\frac{1}{2} \begin{pmatrix} 1+2\,n^{\rm{th}}_-&0&0&0\\0&1+2\,n^{\rm{th}}_-&0&0\\ 0&0&1+2\,n^{\rm{th}}_+&0\\0&0&0&1+2\,n^{\rm{th}}_+ \end{pmatrix}. 
\end{equation}
Note that since we are considering small detuning $$\omega_+\simeq\omega_-\simeq\omega_d/2$$ and then $$n^{\rm{th}}_+\simeq n^{\rm{th}}_-\simeq n^{\rm{th}}.$$ Using Eqs. (\ref{eq:cuadratures}) and (\ref{eq:covariancematrixin}), we obtain the covariance matrix of the outgoing modes  
\begin{eqnarray}\label{eq:covariancematrixout}
V &=&\frac{1}{2} \begin{pmatrix} A & C\\ C^T & B\end{pmatrix}, \nonumber\\
A&=&1+2\,n^{\rm{th}}_- +f^2 (1+2\,n^{\rm{th}}_+)\openone,\nonumber\\ 
B&=&1+2\,n^{\rm{th}}_+ +f^2 (1+2\,n^{\rm{th}}_-)\openone, \nonumber\\
C&=&2f (1+ n^{\rm{th}}_+ + n^{\rm{th}}_-)\sigma_x.
\end{eqnarray}

This is a two-mode squeezed thermal state characterised by the squeezing parameter $2f\,$ and its standard form is obtained by just replacing $\sigma_x$ by $\sigma_z$ in $C$. The main aim of this work is to characterise the quantum correlations of the state described by the covariance matrix in Eqs. (\ref{eq:covariancematrixout}). As a measurement of quantum correlations, we choose quantum discord
which can be exactly computed for two-mode squeezed thermal states, as
recently proven by \cite{pirandola2014}. In fact, the optimality result of Ref.
\cite{pirandola2014} implies that, for these states, the (unrestricted) quantum
discord coincides with the upper bound of the Gaussian discord \cite{giordaparis2010,addesodatta}
which here takes the form::
\begin{equation}\label{eq:gaussianquantumdiscord}
C(V) = h(\sqrt{I_2})-h(d_-)-
h(d_+)+h(\frac{\sqrt{I_1}+2
\sqrt{I_1 I_2} + 2 I_3}{1+2 \sqrt{I_2}}),
\end{equation}
where $$I_1=\det A,\, I_2=\det B,\, I_3=\det C,$$ and $d_+$, $d_-$ are the symplectic eigenvalues, that is the eigenvalues of the matrix $ i\Omega\,V$ -where $\Omega=\left(-i\sigma_y, 0; 0, -i\sigma_y\right)$- and 
\begin{equation}\label{eq:h}
h(x)=(x+\frac12)\log_{2}(x+\frac12)-(x-\frac12)\log_{2}(x-\frac12). 
\end{equation}
Note that the quantum discord is not symmetric under the exchange of modes \cite{giordaparis2010}, and the expression resulting of the exchange of $+$ and $-$ would replace $I_1$ by $I_2$. In our case, however, the only effect of the replacement is the exchange of $n^{\rm{th}}_{+}$ and $n^{\rm{th}}_{-}$, which we are considering to be approximately equal in the regime of small detuning. Therefore, we can consider just a single expression for the quantum discord, Eq. (\ref{eq:gaussianquantumdiscord}). 

Using Eqs. (\ref{eq:covariancematrixout}) and (\ref{eq:gaussianquantumdiscord}), we find that in the perturbative regime: $d_{\pm}=1/2 -f^2/2+n^{\rm{th}}_{\pm}$ and finally
\begin{equation}
C(V) =\rm{max} \{0,f^2-\frac{(n^{\rm{th}})^2}{2}\}.
\end{equation}
Therefore, the outgoing state displays nonzero discord as long as $f>n^{\rm{th}}/\sqrt{2}$. For the sake of comparison with the results of \cite{nonclassicaldce}, we write it explicitly in terms of the driving amplitude $\epsilon$, finding that the onset of quantum discord occurs at:
\begin{equation}
\epsilon_0 = \frac {\sqrt{2}v}{L^0_{\rm eff}\omega_d} n^{\rm th}
\end{equation}
which represents an improvement of $\sqrt{2}$ with respect to the results reported in \cite{nonclassicaldce} for the logarithmic negativity \cite{vidalwerner, plenio} and other non classicality indicators. 
\begin{figure}[h!]
\includegraphics[width=\columnwidth]{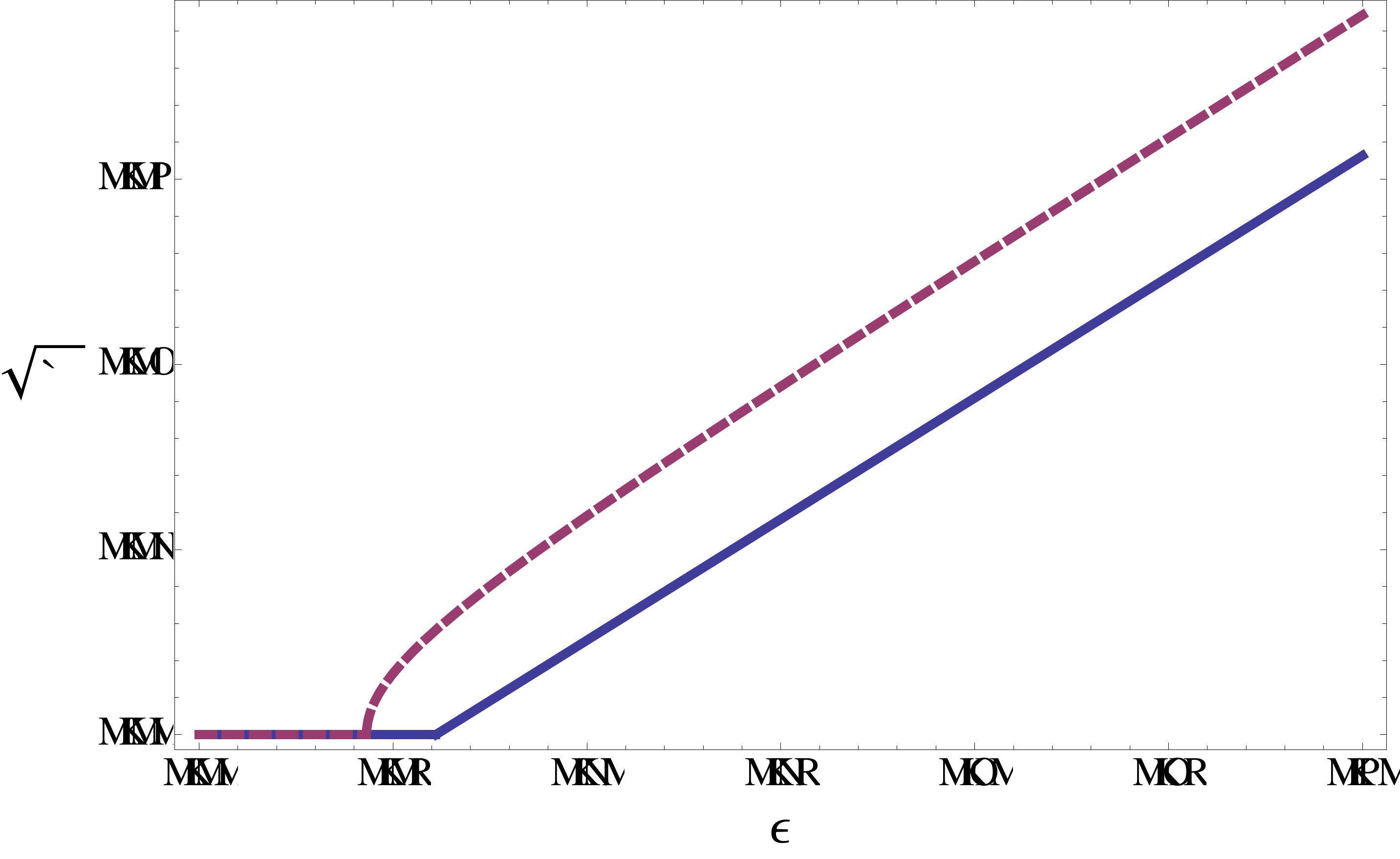}
\caption{\label{fig:fig1} Quantum discord $\sqrt{C}$ -red, dashed- and logarithmic negativity -blue, solid- as a function of the normalised driving ampitude $\epsilon$. We consider experimental parameters $v=1.2\cdot 10^8 \rm{m/s}$, $\omega_d=2\pi\cdot 10\rm{GHz}$, $L_{\rm eff}(0)=0.5\rm{mm}$ and $T=50 \rm mK$. Thus the small parameter $f<0.05$ is well within the perturbative regime, as well as the average numbers of thermal photons $n^{\rm th}\simeq 8\cdot 10^{-3}$. The onset of quantum discord appears before the entanglement one and the magnitude of quantum discord is always larger than entanglement.}
\end{figure}
\begin{figure}[h!] 
\includegraphics[width=\columnwidth]{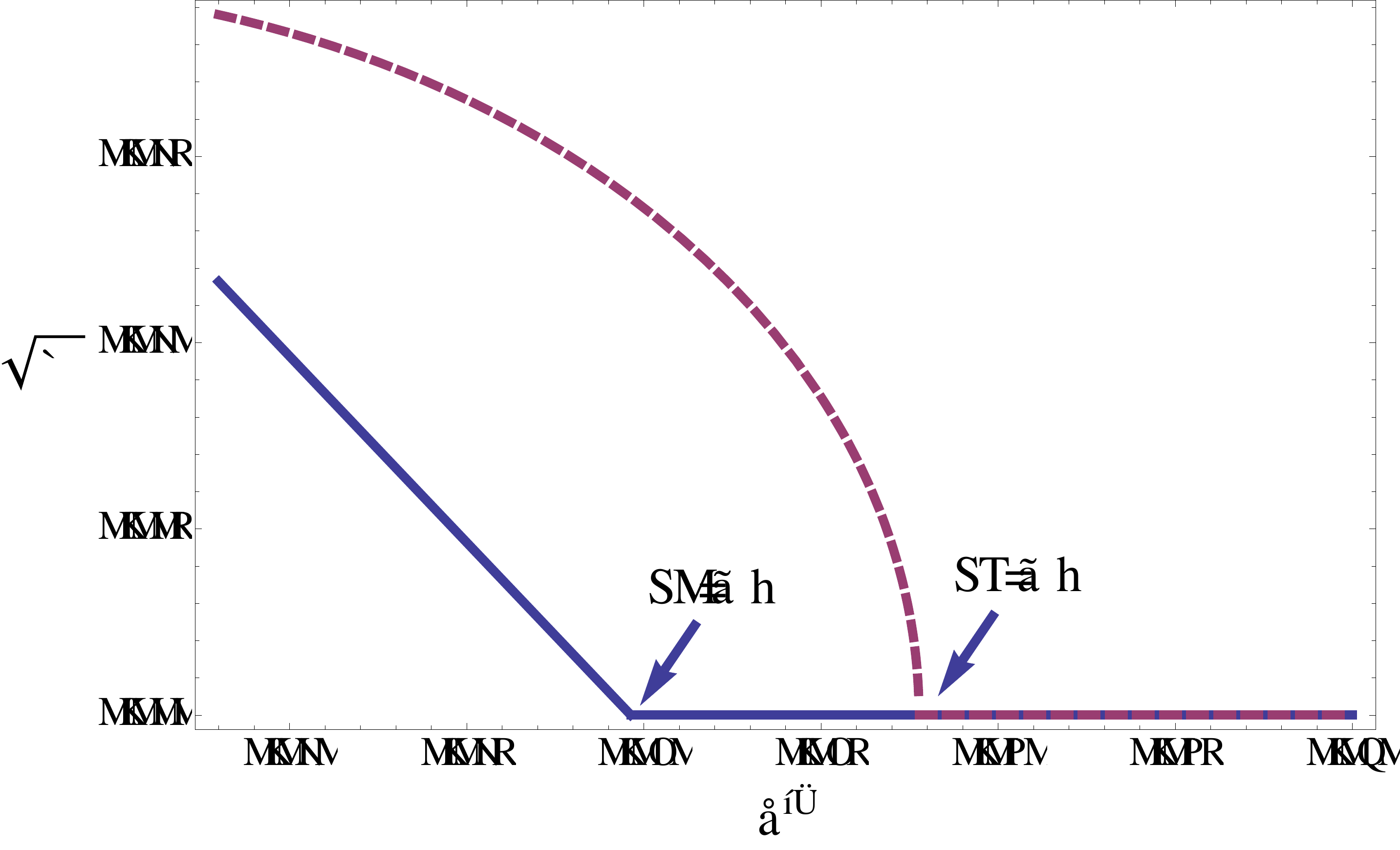}
\caption{\label{fig:fig2}
Quantum discord $\sqrt{C}$ -red, dashed- and logarithmic negativity -blue, solid- as a function of the average number of thermal photons $n^{\rm th}$. We consider experimental parameters $v=1.2\cdot 10^8 \rm{m/s}$, $\omega_d=2\pi\cdot 10\rm{GHz}$, $L_{\rm eff}(0)=0.5\rm{mm}$ and $\epsilon=0.15$. Thus the small parameter $f<0.05$ is well within the perturbative regime. While quantum entanglement vanishes at $T\simeq60\rm mK$,  quantum discord is more robust and survives up to $T\simeq 67 \rm mK$.}
\end{figure}

In Figs. (\ref{fig:fig1}) and (\ref{fig:fig2}) we compare the behaviour of $\sqrt{C}$ with the logarithmic negativity computed in \cite{nonclassicaldce} for realistic experimental parameters. We choose $\sqrt{C}$ instead of $C$ itself in order to include the same perturbative orders in both cases, as explained in \cite{massimofermi}. We find that the amount of quantum discord is always larger than the entanglement and that the conditions for the existence of nonzero correlations are less demanding in the case of discord.  In particular, for a fixed value of the temperature the discord is larger than 0 at values of the driving amplitude where entanglement is still 0 -Fig. (\ref{fig:fig1})- or conversely, for a fixed value of the driving amplitude the discord is still finite at values of the temperature where the entanglement has already vanished -Fig. (\ref{fig:fig2}). Therefore, the experimental conditions required to achieve quantum correlations in the DCE radiation are less demanding than those required to achieve entanglement. This could help the experimental characterisation of the quantum nature of the DCE radiation. Moreover, these results widen the applicability of DCE radiation for quantum technologies. As a first particular example, the states in Eq. (\ref{eq:covariancematrixout}) have proven to be the resource in a device-dependent Quantum Key Distribution protocol in the parameter regime where they possess zero entanglement and non-zero discord (see the Supplementary Information of \cite{discordcryp}).

In summary, we have considered the experimental scenario of the DCE experiment in a superconducting waveguide terminated by a SQUID and analysed the quantum correlations of the generated radiation, as characterised by the quantum discord instead of entanglement. We have found that quantum discord is non-zero at realistic experimental values of the driving amplitude and temperature where entanglement is zero.  Therefore, we have extended the experimental parameter regime where the quantum nature of the DCE radiation in superconducting scenarios can be assessed. Interestingly, in this new parameter regime where discord is different from zero while entanglement remains null we have identified a first technological application in the realm of continuous variable quantum cryptography. A thorough investigation of the role of DCE quantum discord in quantum technologies lies beyond the scope of the current work, but promising possibilities might include quantum interferometric setups \cite{discordtech3} or remote state preparation protocols \cite{discordtech1} where quantum discord is a key ingredient. This would complement current entanglement-based investigations on the use of DCE in quantum technologies \cite{felicetti}. Moreover, our analysis can be extended to different quantum platforms. In particular, the question of the quantum nature of the phononic DCE radiation in Bose-Einstein condensates remains open \cite{casimirwestbrook} so a similar approach as the one taken here should be of interest in that case as well. 

G J acknowledges funding from the Swedish Research Council and the European Research Council.

\end{document}